\documentclass[aps,pra,preprint,groupedaddress]{revtex4-1}
\usepackage{graphicx}% Include figure files
\usepackage{dcolumn}% Align table columns on decimal point
\usepackage{bm}% bold math
\usepackage{csquotes}

\begin{document}

% Use the \preprint command to place your local institutional report
% number in the upper righthand corner of the title page in preprint mode.
% Multiple \preprint commands are allowed.
% Use the 'preprintnumbers' class option to override journal defaults
% to display numbers if necessary
%\preprint{}

\title{Micro-lensed single-mode optical fiber with high numerical aperture}

\author{Shinya Kato, Sho Chonan, and Takao Aoki}
\email[]{takao@waseda.jp}
\affiliation{Department of Applied Physics, Waseda University, 3-4-1 Okubo, Shinjuku, Tokyo 169-8555, Japan}

\begin{abstract}
We show that the output mode of a single-mode optical fiber can be directly focused to a sub-wavelength waist with a finite working distance 
by tapering the fiber to a diameter of the order of the wavelength and terminating it with a spherically/hemispherically shaped tip.
Numerical simulations show that a beam waist with a width of as small as 0.62\,$\lambda$ can be formed.
We fabricate micro-lensed fibers and construct a probe-scanning confocal reflection microscope.
Measurements on gold nano-particles show a spatial profile with a width of 0.29\,$\lambda$ for $\lambda = $ 850 nm, 
which is in good agreement with the numerical simulations.
Due to their monolithic structures, these micro-lensed fibers will be flexible substitutes for conventional compound lenses in various experimental conditions such as cryogenic temperature and ultra-high vacuum. 
\end{abstract}

% insert suggested PACS numbers in braces on next line
% insert suggested keywords - APS authors don't need to do this
%\keywords{}

%\maketitle must follow title, authors, abstract, \pacs, and \keywords
\maketitle

\section{Introduction}
An objective lens with large numerical aperture (NA) is an essential component in various experiments ranging from biomedical imaging with confocal microscopy\,\cite{Pawley:vn} to quantum optics experiments demonstrating single photon sources with nano-emitters such as 
single molecules\,\cite{Lounis:2000he}, quantum dots\,\cite{Imamoglu:2000fq}, nitrogen-vacancy (NV) centers in diamond\,\cite{Kurtsiefer:2000de,Brouri:2000jt}, trapped ions\,\cite{Diedrich:1987gc} or neutral atoms\,\cite{Darquie:2005fr}.
In such experiments, high-NA objectives which consist of multiple optical elements are used to obtain high-resolution images or to achieve high photon collection efficiency. 
However, such compound lens systems cause experimental complexity in particular when the systems are used in the extreme conditions, e.g, 
ultra-low cryogenic temperature to suppress thermal contribution from the ambient environment or ultra-high vacuum condition for stable trapping of ions or neutral atoms.

In the applications mentioned above, a single-mode optical fiber often plays an important role as an extremely low-loss waveguide and/or an ideal spatial filter.
It is naturally required to achieve high coupling efficiency of photons into the single-mode optical fibers from observation objects in the microscopy\cite{Kimura_Appl_Opt_1991} or from nano-emitters.
However, additional coupling systems inevitably increase the experimental complexity and optical losses, and reduce system robustness and stability.
Therefore, it is desirable to develop monolithic lens-fiber systems, i.e., lensed fibers, with high NA. 
Various types of lensed fibers have been developed, including lens structures fabricated on the facets of standard single-mode fibers\cite{Cohen_Appl_Opt_1974,Bachelot_Appl_Opt_2001,Eah_RSI_2003} and tapered fibers with relatively large diameter (20\,$\mu$m) with hemispherical end tip \cite{Kuwahara_Appl_Opt_1980}. These conventional lensed fibers are used for directly coupling outputs of diode lasers into single-mode fibers 
and they have relatively small numerical apertures or large losses.

In this paper, we report on micro-lensed single-mode optical fibers with high NA and low losses. Standard single-mode optical fibers are tapered to diameters of the order of the wavelength and are terminated with a spherically/hemispherically shaped tips. 
Numerical simulations show that the output modes of the micro-lensed fibers are focused to sub-wavelength waists with finite working distances.
In addition, we experimentally demonstrate confocal microscopy using micro-lensed fibers as a high-NA objective.
The obtained image shows good agreement with the numerical simulation.
A significant advantage of the demonstrated micro-lensed fibers is the direct coupling of the collected photons into the single mode fiber.
This enables one to perform efficient spatial filtering and low-loss long-distance transmission of the collected photons.
The monolithic connection of a high-NA lens and a single-mode optical fiber can remarkably reduce the complexity and enhance the stability 
of the experiments in the extreme conditions, and they can be easily integrated into fiber optical systems.

\section{Design of lensed fiber tips and FDTD simulations}

It has been reported that the guided mode of a standard single-mode fiber can be transferred to the fundamental mode of an air-clad micro-/nano-fiber 
through an adiabatically tapered region \cite{Love:1986, Birks:1992ks, Aoki:2010}. Therefore, if it is possible to focus the output of the fundamental mode of the micro-/nano-fiber to a sub-wavelength waist by appropriately shaping the tip of the micro-/nano-fiber, the whole structure of the taper, micro-/nano-fiber, and the shaped tip functions as a monolithic, high-numerical-aperture lens for 
the single-mode fiber. 

%%%%%%%%%%%%%%%%%%%%%%%%%%%%%%%%%%%%%%%%%%%%%%%%%%%%%%%%
\begin{figure}[htbp]
   \centering
   \includegraphics[width=10 cm]{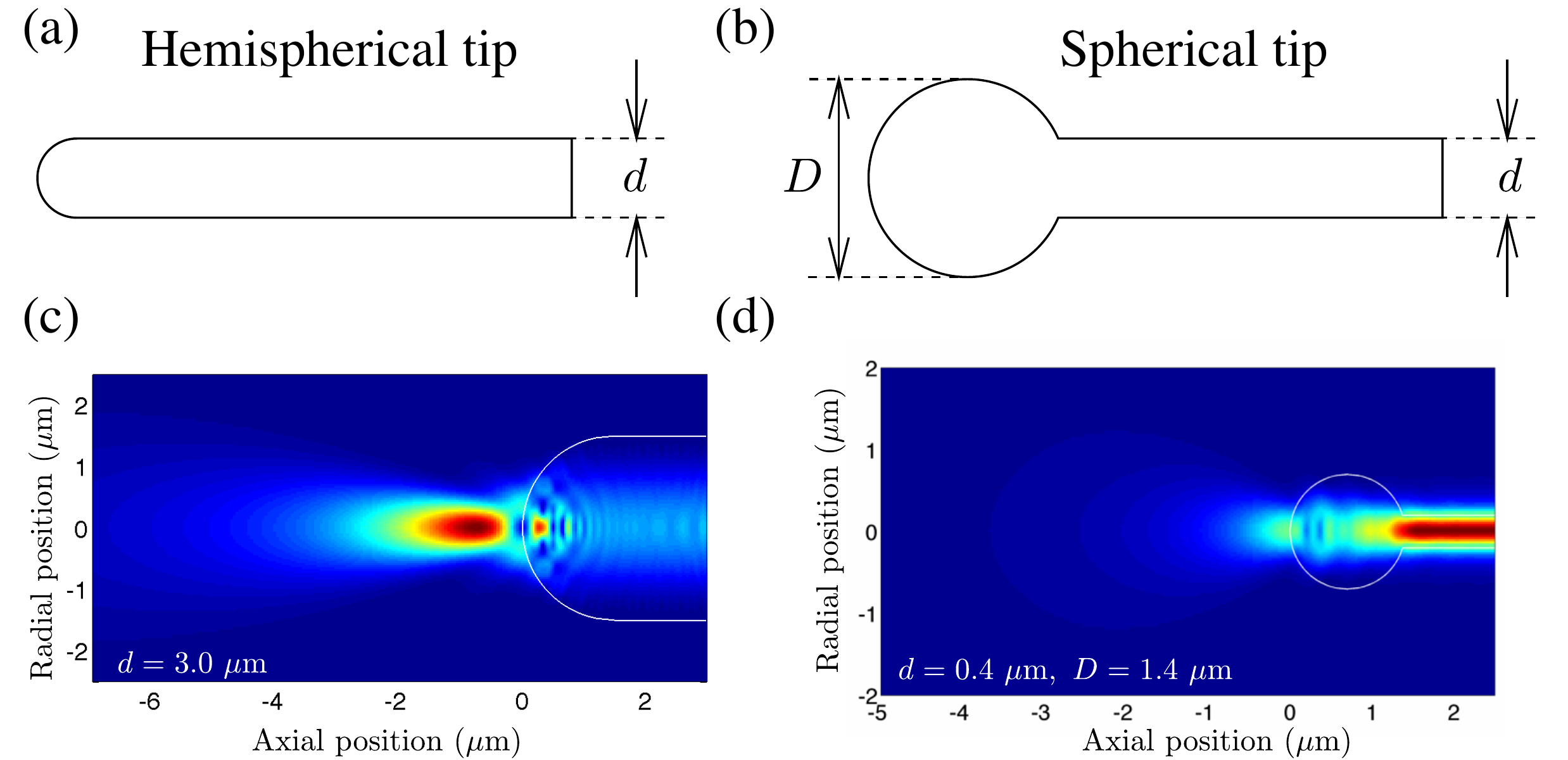}
   \caption{Schematics of fibers with (a) the hemispherical and (b) the spherical tips.
   $d$ and $D$ are the diameters of the stem fiber and the tip sphere, respectively ($d=D$ for the hemisphere case).
   Numerically calculated intensity distribution of the output beam from (c) the hemispherical and (d) the spherical tips.}
   \label{fig:sketch}
\end{figure}
%%%%%%%%%%%%%%%%%%%%%%%%%%%%%%%%%%%%%%%%%%%%%%%%%%%%%%%%

First we investigate the mode profiles of the outputs of air-clad silica micro-/nano-fibers with two types of tips, hemispherical and spherical, as shown in Fig.\,\ref{fig:sketch}(a) and (b).
We perform numerical simulations based on three-dimensional finite-difference-time-domain (FDTD) method\,\cite{fdtd}.
In the simulations, we use the numerically solved fundamental propagation mode in the fiber as the input light source.
We assume that the wavelength of the light source is 850 nm throughout this paper. 
Correspondingly, we assume the refractive index of silica $n = 1.45$.
Note that any wavelength-dependence in the present simulations originates from diffraction and material dispersion. 
In the absence of material dispersion, the calculation result would be independent of the wavelength when the length is normalized with the wavelength.

\begin{figure}[htbp]
   \centering
   \includegraphics[width=13 cm]{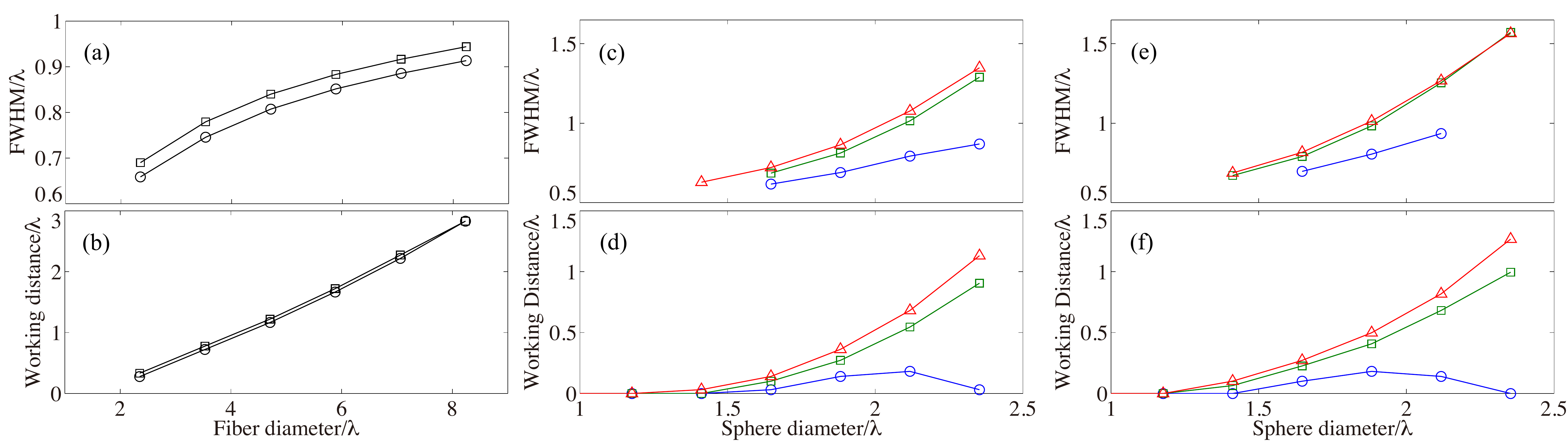}
   \caption{
   (a) FWHM and (b) WD of the output beam waist as functions of the fiber diameter $d$ with a hemispherical tip.
   Circles and squares correspond to the results on the plane parallel and perpendicular to the polarization of the propagating mode of the light, respectively.
   (c)-(f) FWHM and WD of the output beam waist as functions of the sphere diameter $D$ with a spherical tip. 
   Circles, squares, and triangles correspond to the fiber diameter $d$=400, 500, 600 nm, respectively.
   (c) and (d) are the results on the plane parallel to the polarization, while (e) and (f) are those on the plane perpendicular to the polarization.
   }
   \label{fig:sim}
\end{figure}

Typical calculation results of the intensity distributions of the output beams are shown in Fig.\,\ref{fig:sketch}(c) and (d).
It can be clearly seen that the beams are tightly focused with small but finite working distances from the tips of the fibers. 
We examine the full width at half maximum (FWHM) of the beam waist and the working distance (WD) for various values of $d$ and $D$.
The results are shown in Fig.\,\ref{fig:sim}.

%%%%%%%%%%%%%%%%%%%%%%%%%%%%%%%%%%%%%%%%%%%%%%%%%%%%%%%%
%%%%%%%%%%%%%%%%%%%%%%%%%%%%%%%%%%%%%%%%%%%%%%%%%%%%%%%%

As for the hemispherical tips, finite working distances (WD $>0$) are obtained for $d>2.4\, \lambda$.
The smallest FWHM of the waist is 0.66\,$\lambda$.
It should be noted that the single-mode cutoff diameter of air-clad fiber with $n=1.45$ is $d_c = 0.73 \lambda$. 
Therefore the air-clad fibers with $d>2.4 \lambda$ guide multiple modes.
However, photons collected into the fundamental mode of the air-clad fiber from a nano-emitter located at the mode waist 
are efficiently guided into the standard silica-clad single-mode fiber through the adiabatically tapered region\,\cite{Chonan}.

When the tips are spherical, on the other hand, single-mode air-clad fibers can have finite working distances as shown in 
Fig.\,\ref{fig:sim}(c)-(f). The smallest FWHM of the waist is 0.62\,$\lambda$ which corresponds to the NA of 0.85. 

\section{Fabrication and characterization of lensed fiber tips}
\begin{figure}[htbp]
   \centering
   \includegraphics[width=9 cm]{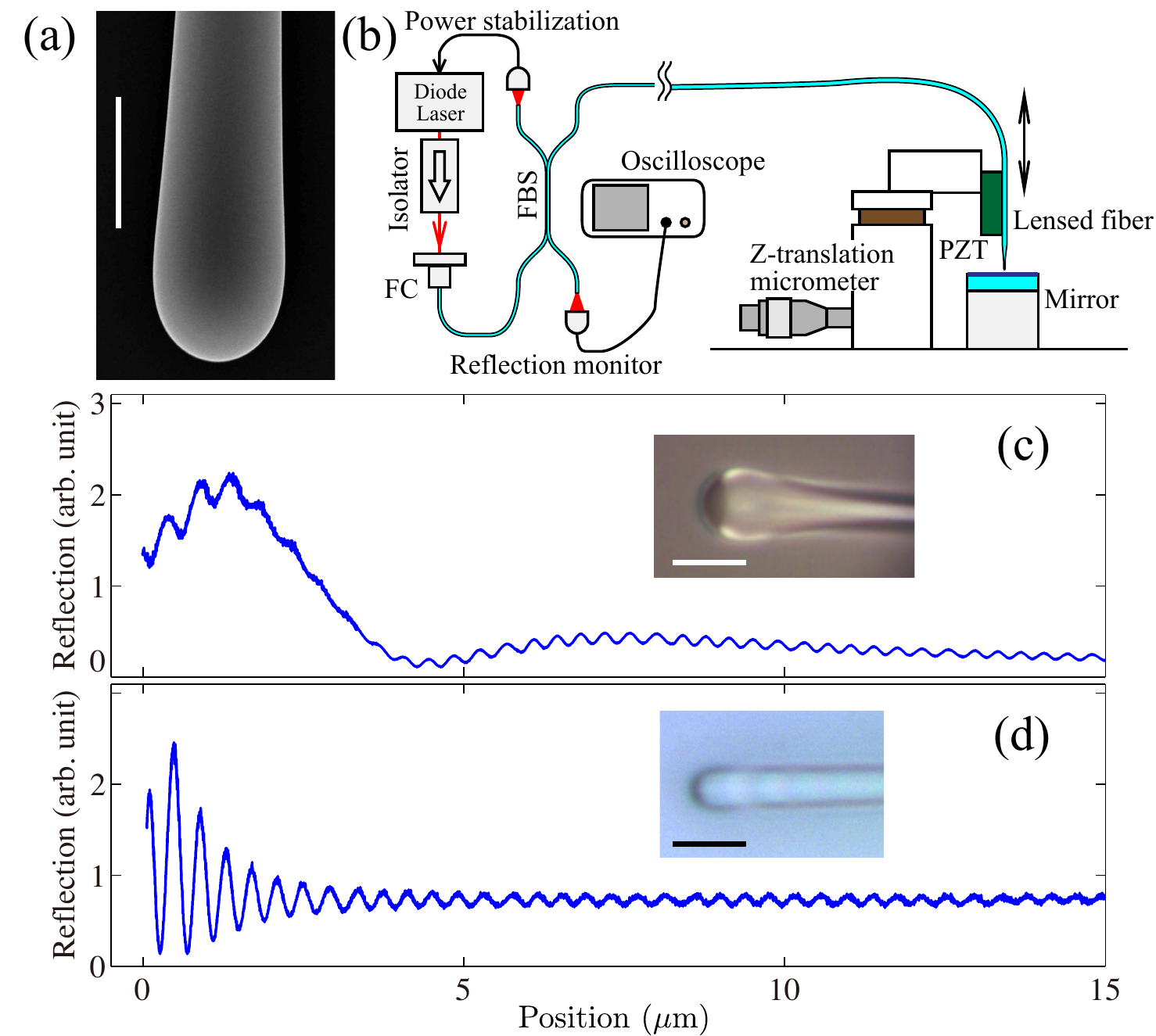}
   \caption{(a) SEM image of the hemispherical fiber tip.
   The scale bar corresponds to 3 $\mu$m.
   (b) Schematic diagram of the experimental setup.
   FC and FBS denote a fiber coupler and a fiber beam splitter, respectively.
   (c)-(d) Measured reflection beam power as a function of the distance between the tip and the mirror surface.
   Insets in (c)-(d) are the optical microscope images of the fiber tips.
   The scale bars correspond to 10 $\mu$m and 5 $\mu$m in the insets of (c) and (d), respectively.}
   \label{fig:exp-1}
\end{figure}

Next we fabricate micro-lensed fibers. 
We start from drawing a silica fiber taper with the flame-brush method\,\cite{Birks:1992ks, Aoki:2010}.
A computer-controlled rig with a hydrogen-oxygen torch heats and pulls a standard single-mode optical fiber (Thorlabs, SM830-5.6-125) with 125 $\mu$m clad until the fiber is tapered to a desired diameter. % which is down to 1 $\mu$m in this work.
After the heat-and-pull procedure, we cut the fiber taper in half.
Then we anneal the tip of the taper with the hydrogen-oxygen torch to form the hemispherical shape.
In Fig.\,\ref{fig:exp-1}(a), we show a typical scanning electron microscope (SEM) image of a hemispherical lensed fiber fabricated by this procedure.

%%%%%%%%%%%%%%%%%%%%%%%%%%%%%%%%%%%%%%%%%%%%%%%%%%%%%%%%
%%%%%%%%%%%%%%%%%%%%%%%%%%%%%%%%%%%%%%%%%%%%%%%%%%%%%%%%

In order to characterize the output mode of the hemispherical lensed fiber, we perform reflection measurements as schematically shown in Fig.\,\ref{fig:exp-1}(b).
We use a diode laser operating at 850 nm as a light source.
The output of the laser is coupled into a single-mode optical fiber, and split by a 50:50 optical fiber beam splitter (FBS).
One of the two output ports is used as the power monitor for stabilization of the laser power.
The other port is fusion-spliced to the micro-lensed fiber.
Since the working distance and the waist size of our lensed fiber is of the order of the wavelength, sub-wavelength position resolution 
is required. The vertically-aligned lensed fiber tip is fixed on a piezo transducer to achieve high-resolution positioning. 
A micrometer-driven z-translation stage is used for coarse positioning.
As we scan the distance between the fiber tip and the mirror surface, we measure the intensity of the light coupled back to the fiber 
through the lensed tip after reflecting on the mirror surface.

The measured retro-reflected beam powers as functions of the tip-mirror distance are plotted in Fig.\,\ref{fig:exp-1}(c) and (d) 
for the fiber diameters $d$ of 11 $\mu$m and 3 $\mu$m, respectively.
Sinusoidal oscillations are observed due to interference between the reflection at the tip and that at the mirror surface.
The envelope of the reflection intensity has its maxima at finite tip-mirror distances, which clearly indicates focusing of 
the output beam from the tip. Note that the position of reflection maxima depends on the fiber diameter.

\section{Confocal microscopy with lensed fiber tips as objectives}
Lastly, we demonstrate probe-scanning confocal microscopy using the hemispherical micro-lensed fiber. 
We use a standard positive grid array of chrome on a glass slide (Thorlabs, R1L3S3P) and gold nano-particles (GNP) with a mean dimeter of 100 nm (Sigma-Aldrich, 742031) dispersed on a glass slide as observation objects.
The coefficient of variance of the GNP size distribution is less than 12 \%.
The experimental setup is almost the same as the one described in the previous section, but the fixed mirror is replaced by the objects fixed on a xy-stage with piezo actuators to scan on the horizontal plane.
We use two hemispherical micro-lensed fibers with the tip diameter of 7 $\mu$m and 2.6 $\mu$m. 
As for the 7-$\mu$m tip case, we actively stabilize the vertical position of the tip by keeping the reflected beam intensity at an interference peak with feedback to the piezo element on which the fiber is fixed.
As for the 2.6-$\mu$m tip case, on the other hand, the fiber is supported by a solid plate on the piezo element to suppress vertical position fluctuation without active stabilization. 
The normalized reflection signals are measured as functions of the horizontal position of the observation object, which is derived from the applied voltage on the piezo actuator for the horizontal scan. 

%%%%%%%%%%%%%%%%%%%%%%%%%%%%%%%%%%%%%%%%%%%%%%%%%%%%%%%%
\begin{figure}[htbp]
   \centering
   \includegraphics[width=10 cm]{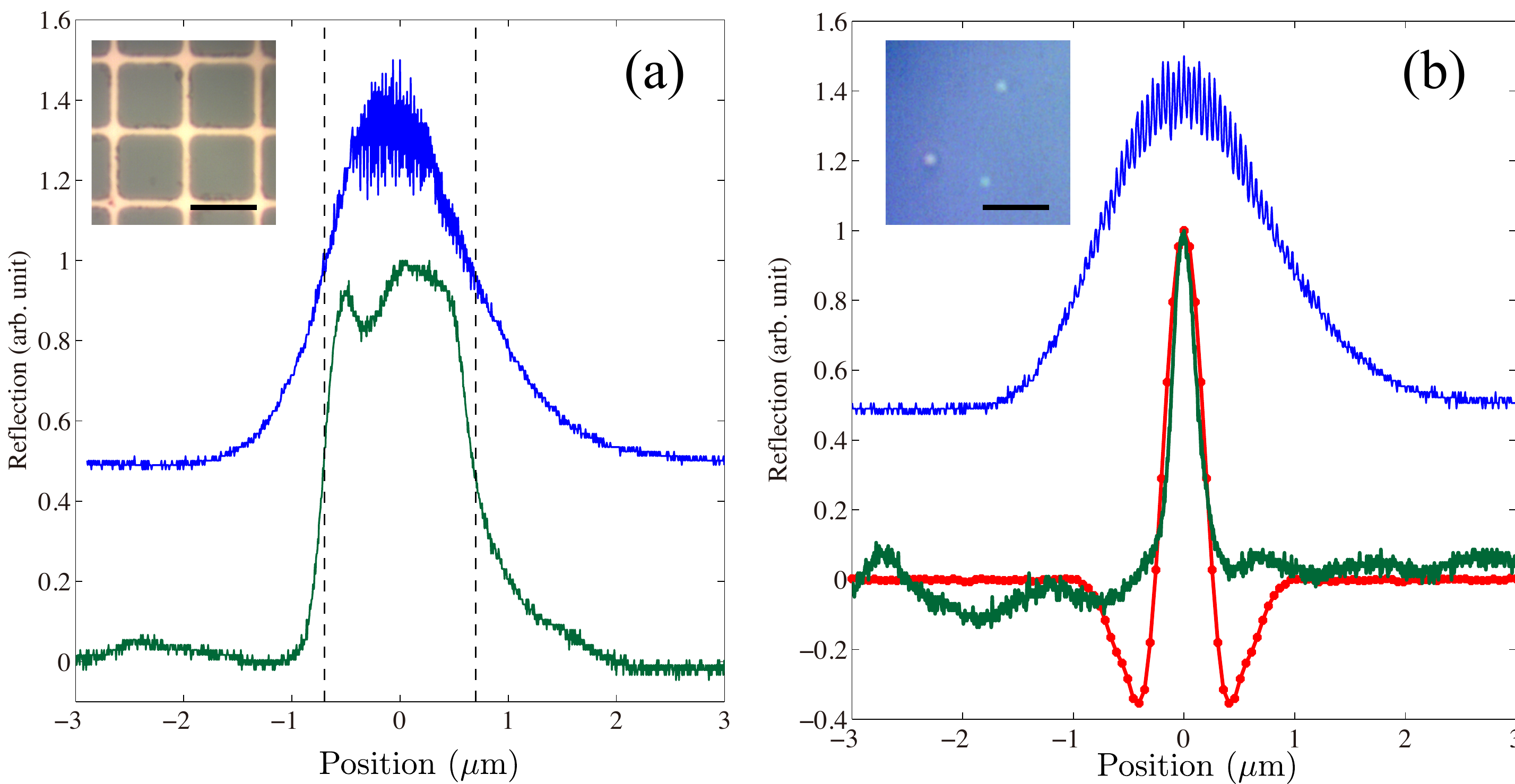}
   \caption{The obtained profile of (a) a Cr line in a grid array and (b) a GNP in the confocal microscopy with hemispherical fiber tips.
   Insets in (a) and (b) display conventional optical microscope images of the Cr grid and the GNPs.
   The scale bars in the insets correspond to 10 $\mu$m and 4 $\mu$m for (a)  and (b), respectively.
   The blue and green line data are taken with the fiber tip whose diameters are 7 $\mu$m and 2.6 $\mu$m, respectively.
   These lines are vertically offset for clarity.
   The dashed lines in (a) denote the Cr stripe width of 1.4 $\mu$m which was determined by the optical microscope image. 
   The red circles in (b) denote the simulation result. 
   }
   \label{fig:exp-2}
\end{figure}
%%%%%%%%%%%%%%%%%%%%%%%%%%%%%%%%%%%%%%%%%%%%%%%%%%%%%%%%

Figures \ref{fig:exp-2} (a) and (b) show the obtained cross section profiles for a Cr line and a GNP, respectively.
The step-like profile of the Cr line is well resolved with the 2.6-$\mu$m tip, while it is not clearly resolved with the 7-$\mu$m tip.
The difference of the resolution of the two fiber tips is more obvious for the measurements on the GNP shown in Fig.\,\ref{fig:exp-2}(b).
The FWHM of the profile obtained with 2.6-$\mu$m tip is as narrow as  250\,nm (0.29\,$\lambda$). 
Note that the resolution is in the sub-diffraction-limited regime.

We perform a FDTD simulation modeling the present confocal microscopy to compare with this observation.
In the simulation, we consider a gold sphere with a diameter of 100 nm on a silica slide and a 2.6-$\mu$m-diameter silica fiber with a hemispherical tip 
whose axis is normal to the slide surface.
The fundamental mode at the wavelength of 850 nm in the fiber is used to simulate the input beam. 
The optical power guided in the fundamental mode propagating to the backward direction is calculated while changing the horizontal position of the fiber tip relative to the GNP with the vertical position of the tip kept constant at 400 nm from the silica slide.
The simulation result is also displayed in Fig.\,\ref{fig:exp-2}, which show good agreement with the experimental observation.

We attribute the observed sub-diffraction-limited resolution to the interference between the beam reflected at the observation object and 
that reflected at the fiber tip.
It has been reported that sub-diffraction-limited resolution can be observed as an artifact due to topological effects 
in reflection scanning near-field optical microscopy with uncoated fiber tips\cite{Sandoghdar:1997fu}, in which the 
interference between two reflected beams plays an important role.
On the other hand, the sub-diffraction-limited resolution observed in the present study does not originate from topological effects. 
In fact, the probe height is kept constant in the present measurement, eliminating any topological effect.

\section{Conclusion}
In conclusion, we investigated micro-lensed single-mode optical fibers with spherically/hemispherically shaped tips.
The numerical simulations showed focusing of the output modes for both shapes.
Furthermore, hemispherical fiber tips were fabricated and a probe-scanning confocal microscope was constructed.
The observed cross-sectional profile of the GNP was in good agreement with the numerical simulation.
The presented micro-lensed fiber has various advantages compared to the conventional high-NA compound lens systems.
One of the most significant advantages is the direct coupling to a single-mode optical fiber.
This enables one to transmit photons collected through the lensed fiber over a long distance with low loss, 
as well as to combine outputs from two different microscope systems with a perfect spatial mode matching, 
which is crucial for quantum information protocols using indistinguishable single photons.
In addition, the simple and robust structure of the micro-lensed fiber is suited for various experimental conditions 
such as the cryogenic temperature or the ultra-high vacuum.

\begin{acknowledgments}
TA thanks Mr. M. Kawaguchi for assistance at the preliminary stage of this work. 
This work is partially supported by SCOPE and MEXT KAKENHI Grant Number 23104717.
\end{acknowledgments}

% Create the reference section using BibTeX:
%\bibliography{basename of .bib file}

\end{document}